\documentclass[useAMS,usenatbib]{mn2e}
\usepackage[utf8]{inputenc}
\usepackage{amsmath}
\usepackage{amssymb}
\usepackage{amsbsy}
\usepackage{graphicx}
\usepackage{units}
\usepackage{url}
\usepackage{color}

\newcommand{\Bbold}{\bmath{B}}

\newcommand{\tff}{\,t_{\text{ff}}}
\newcommand{\msun}{\text{M}_{\sun}}
\newcommand{\mug}{\umu\unit{G}}
\newcommand{\alfven}{Alfv\'{e}n}

\begin{document}

\title[Protostellar jets and outflows with SPMHD]{Protostellar outflows with Smoothed Particle Magnetohydrodynamics (SPMHD)}

\author{Florian Bürzle}

\author[F. Bürzle et al.]{Florian Bürzle$^{1}$\thanks{E-mail: florian.buerzle@uni-konstanz.de}, Paul C. Clark$^ {3}$, Federico Stasyszyn$^{2}$, Klaus Dolag$^{2}$ \newauthor and Ralf S. Klessen$^{3}$ 
 \newauthor\\
$^{1}$Universität Konstanz, Fachbereich Physik, Universitätsstr. 10, 78464 Konstanz, Germany\\
$^{2}$Universitütssternwarte München, Scheinerstr. 1, 81679 München, Germany\\
$^{3}$Zentrum für Astronomie der Universität Heidelberg, Institut für Theoretische Astrophysik,\\ Albert-Ueberle-Str. 2, 69120 Heidelberg, Germany}

\pagerange{\pageref{firstpage}--\pageref{lastpage}} \pubyear{2011}

\maketitle

\label{firstpage}
\bibliographystyle{mn2e}

\begin{abstract}
The protostellar collapse of a molecular cloud core is usually accompanied by outflow phenomena. The latter are thought to be driven by magnetorotational processes from the central parts of the protostellar disc. While several 3D AMR/nested grid studies of outflow phenomena in collapsing magnetically supercritical dense cores have been reported in the literature, so far no such simulation has been performed using the Smoothed Particle Hydrodynamics (SPH) method. This is mainly due to intrinsic numerical difficulties in handling magnetohydrodynamics within SPH, which only recently were partly resolved. In this work, we use an approach where we evolve the magnetic field via the induction equation, augmented with stability correction and divergence cleaning schemes. We consider the collapse of a rotating core of one solar mass, threaded by a weak magnetic field initially parallel to the rotation axis so that the core is magnetically supercritical. We show, that Smoothed Particle Magnetohydrodynamics (SPMHD) is able to handle the magnetorotational processes connected with outflow phenomena, and to produce meaningful results which are in good agreement with findings reported in the literature. Especially, our numerical scheme allows for a quantitative analysis of the evolution of the ratio of the toroidal to the poloidal magnetic field, which we performed in this work.
\end{abstract}

\begin{keywords}
 ISM: clouds - ISM: jets and outflows  - ISM: magnetic fields - MHD - stars: formation 
\end{keywords}

\section{INTRODUCTION}
Observations in star forming regions have revealed that the protostellar collapse of a molecular cloud core is usually accompanied by outflow phenomena \citep[e. g.][]{Wu2004uq,Bally2007fk}, which were already discovered back in the 1970s \citep{Zuckerman1975fj,Kwan1976yq,Zuckerman1976kx}. Usually, protostellar outflows are classified into two types, optical jets or molecular outflows. The latter typically exhibit slow velocities and wide opening angles, and are observationally identified by line emission from their CO molecules. The former, having narrow opening angles and high velocities, are observed optically. Both phenomena may occur together, where usually the high-velocity jet is enclosed in a low-velocity molecular outflow. The physical mechanisms that drive outflows are still not fully understood. The adiabatic (or 'first') core, the circumstellar disc around the protostar and the protostar itself are possible origins of either jet or molecular outflow or both. First analytical work by \cite{Blandford1982fk} and \cite{Pudritz1983uq} suggested rapid rotating, magnetised protostellar discs as possible origins of these outflows, since other sources like pressure due to either the gas or radiation, have been ruled out since those mechanisms fail to provide sufficient energy and momentum. Later,  \cite{Lynden-Bell1996fk,Lynden-Bell2003fk} introduced the idea of a 'magnetic tower', an outflow driven by magnetic pressure originating in toroidal magnetic field components. 

From a numerical point of view, the simulation by \cite{Tomisaka1998fk} was one of the first to produce a magnetically driven outflow during the collapse of a magnetised and rotating molecular cloud core. There it was shown that the outflow is launched by dynamically building up of a toroidal magnetic field. Other simulations concentrating on outflow phenomena were performed by \cite{Tomisaka2000fj,Tomisaka2002kx}, and \cite{Banerjee2006uq} who followed the evolution up to protostar formation using an ideal MHD approximation and appropriate cooling models. These works in general agree in finding two distinct flows emerging from different regions, namely the low-velocity flow from the first, the high-velocity flow from the second core which, respectively, can be identified as molecular outflow and optical jet. Another work by \cite{Hennebelle2008uq} used also ideal MHD and a barotropic equation of state and found a magnetic outflow best described by a magnetic tower for a model with low magnetic fields strength, and a morphological different flow for a high field strength model with characteristics of the disc-wind model proposed by \cite{Blandford1982fk}. However, their barotropic equation of state only had a prescription for the formation of the first core, and so they were unable to model the evolution beyond this point in the collapse. A similar work by \cite{Machida2005fk}, however starting from a filamentary cloud with initial density perturbations but also using a simple barotropic equation of state, also reported a low-velocity outflow from the first core. \cite{Machida2006uq} used a more sophisticated equation of state that approximates the thermal evolution better, in particular it allows to follow the collapse until second core formation. Furthermore, they used ideal and resistive MHD models, and found a high-velocity jet ejected from the second core in both models. In a following and more detailed study, \cite{Machida2008fk} also stressed the importance of non-ideal MHD in high density regions. They considered models with different initial angular velocities and also found a low-velocity flow from the first core and high-velocity flow from the protostar. However, these authors suggest the former flow being driven by disc-wind mechanisms and the latter by magnetic pressure, conversely to other authors. \cite{Machida2010kc} used again a simple equation of state as in \cite{Machida2005fk}, but in combination with resistive MHD, and applied this scheme to a spherical cloud with initial density perturbations. They concentrated, however, on the formation of the circumstellar disc. They found an outflow driven by the circumstellar disc which, however, is weakened in the later main accretion phase since the amount of infalling gas decreases over time. Finally, the outflow is found to disappear. In a more recent study, which considered the combined effects of MHD and radiation and follows the collapse until first core formation, \cite{Tomida2010fk} found a larger size of the first core and higher entropy in its outer layers, compared to MHD simulations using a barotropic approximation. Furthermore, they find a two-component outflow driven by different mechanisms, namely an inner magnetic pressure mode and an outer magneto-centrifugal mode. 
In this work, we perform SPMHD simulations using a simple set-up with initial parameters comparable to those chosen by \cite{Hennebelle2008uq} and we concentrate on a weak field scenario. We show, that our results agree qualitatively well to those obtained by these authors. In particular, \cite{Hennebelle2008uq} also suggest toroidal magnetic pressure as driver of the low-velocity outflow, in agreement with our results.

\begin{figure}
\begin{flushleft}
\includegraphics[width=\columnwidth]{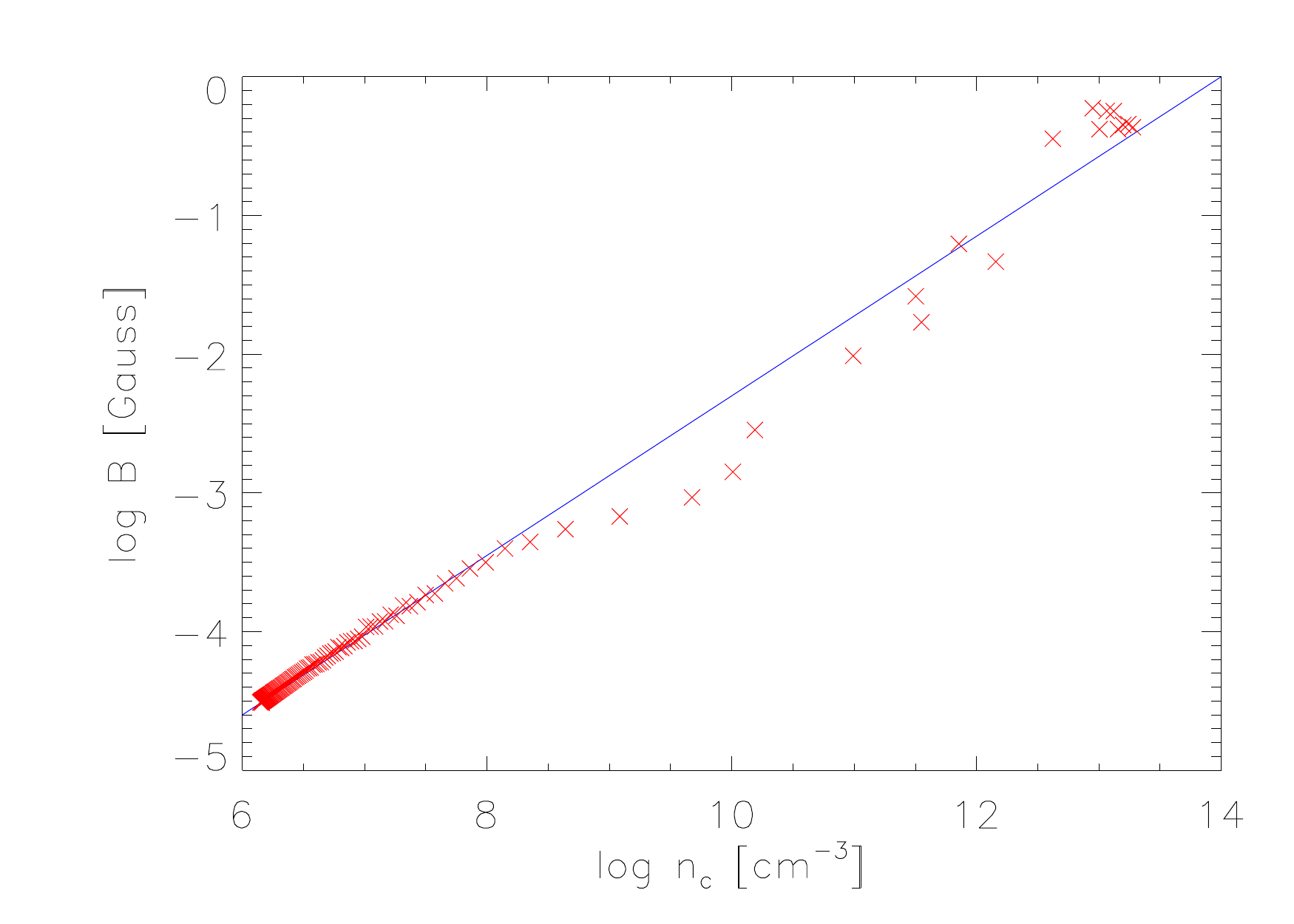}
\caption{\label{fig:ncore_bfield} Magnetic field strength $B$ as a function of the central number density $n_c$ in the core. Each data point represents a spatial average from simulation snapshot which were taken in intervals of $0.01\,\tff$. The solid line was obtained from the power law  $B\propto n^{\kappa}$ with $\kappa = 0.58$.}
\end{flushleft}
\end{figure}

\section{NUMERICAL METHOD AND INITIAL CONDITIONS}
We perform three dimensional simulations using the Smoothed Particle Hydrodynamics (SPH) code \textsc{GADGET} \citep{Springel2001il,Springel2005tw}, which was recently extended by \cite{Dolag2009uq} to allow for the treatment of ideal magnetohydrodynamics. We refer to the latter publication and references therein for a detailed discussion of the algorithms and numerical tests. This implementation of Smoothed Particle Magnetohydrodynamics (SPMHD) has been successfully applied in extra-galactic \citep{Bonafede2011ab} and galactic astrophysics \citep{Kotarba2009fk,Kotarba2010ab,Kotarba2011fk}, as well as to collapse and fragmentation problems in star formation \citep{Burzle2011fk}. All parameters used to control the SPMHD scheme have the same values as adopted in \cite{Burzle2011fk}, except that $\alpha=1$ in this work.

In this work, we use initial conditions reproduced from those given in the work by \cite{Hennebelle2008uq}. The initial cloud mass was chosen to be $M_0=1\,\text{M}_{\odot}$, the radius was set to $R_0 = 4.6\cdot 10^{16}\,\unit{cm}\, \approx 0.015\, \unit{pc}$. This choice corresponds to an initial density of $\rho_0  = 4.87\cdot 10^{-18}\, \unit{g}\,\unit{cm}^{-3}$ and a free-fall time $t_{\text{ff}}\approx 30000\,\unit{yr}$. The spherical cloud is embedded into a low density medium with $\delta=\rho_0/\rho_{\text{ext}}=100$. This medium, however, is not in pressure equilibrium with the cloud ($P_0/P_{\text{ext}}=100$) thus allowing the outer cloud layers to expand into the medium, as in \cite{Hennebelle2008uq}. The whole system is placed into a cube with side length of $4 R_0$ and periodic boundary conditions.

The thermodynamics is controlled by a barotropic equation of state given by
\begin{equation}
P = c_{\text{s,0}}^2\rho \left[ 1 +  \left( \dfrac{\rho}{\rho_{\text{crit}}} \right)^{4/3} \right]^{1/2}
\label{eq:eos}
\end{equation}
where $c_{\text{s},0}=0.2\, \unit{km}\,\unit{s}^{-1}$ is the speed of sound, resulting from a temperature $T=11\,\unit{K}$ and a mean molecular weight of $\umu_{\text{mol}} = 2.1$. The critical density is chosen to be  $\rho_{\text{crit}}=10^{-13}\,\unit{g}\,\unit{cm}^{-3}$, leading, as a consequence of the longer remaining of the gas in the quasi-isothermal regime, to a very rapid growth of the central density. Thus it is difficult to follow the evolution for a long period of time, especially the second collapse is not within the range accessible in this work. Usually this problem can be circumvented by using sink particles \citep{Bate1995jh,Jappsen2005pz,Federrath2010fk}, but since we investigate outflow phenomena launched in the proximity of the 'protostar' we avoid using them here to reduce perturbations to the flow behaviour as far as possible. 

We define the ratio of the mass-to-flux ratio to the critical mass-to-flux ratio as $\mu = (M/\Phi)/(M/\Phi)_{\text{crit}}$, where $\Phi$ is the magnetic flux. For a spherical configuration, the critical value is given by $(M/\Phi)_{\text{crit}} \approx 0.125\, \text{G}^{-1/2}$ \citep{Mouschovias1976dz}. Choosing $\mu=20$ the initial magnetic field strength is given by $30.7\,\mug$, and the plasma beta is given by $\beta_{\text{plasma}} = 56$. The {\alfven} speed is $v_A=B_0 / \sqrt{4 \pi \rho_0}= 3.9 \times 10^{-2}\,\unit{km}\,\unit{s}^{-1}$.

The initial angular velocity is $\Omega_0 = 4.3\cdot 10^{-13}\,\unit{s}^{-1}$ for the cloud which is in solid body rotation, corresponding to $\Omega_0 \tff = 0.4$. Using these choices, we get for the energy ratios $\alpha_{\text{therm}} = E_{\text{therm}}/\left| E_{\text{grav}}\right| = 0.37$ and $\beta_{\text{rot}} = E_{\text{rot}}/\left| E_{\text{grav}}\right| = 0.045$.
\cite{Machida2008fk} characterize their models by a dimensionless parameter $\omega = \Omega_0/\sqrt{4\pi G \rho_0}$ resulting for our choice of initial conditions in a value of $\omega=0.2$, thus comparable with the rapid rotating model considered by these authors.  

The Jeans condition for SPH given by \cite{Bate1997sj} requires that the mass of the cloud needs to be sufficiently resolved, and that the particle softening length needs to be equal to the smoothing length, in order to avoid numerical artefacts during cloud collapse. They proved, that the minimum number of particles required to resolve the cloud mass is given by $N_{\text{min}} = 2 M_0 N_{\text{neigh}}/M_{\text{J}}$, where $M_{\text{J}} $ is the Jeans mass. There exist several formulations of the Jeans mass which differ, as was pointed out by \cite{Nelson2006fk}, by numerical prefactors. We use the most conservative formulation that gives the lowest value for the Jeans mass:
\begin{equation}
M_{\text{J}} = \left(\dfrac{3}{4\pi}\right)^{1/2} \left( \dfrac{5}{2} \right)^{3/2}  \dfrac{c_{\text{s}}^3}{\text{G}^{3/2} \rho^{1/2}}.
\end{equation}
For $c_{\textrm{s}}^3/\rho^{-1/2}$ at $\rho_{\text{crit}}$ we get, using eq. (\ref{eq:eos}), $2^{3/4} c_{\textrm{s},0}^3\rho_{\text{crit}}^{-1/2}$ and thus a value of $N_{\text{min}} \sim 48000$ for $N_{\text{neigh}} = 64$. Employing $N=4 \cdot 10^6$ particles within the cloud ($6 \cdot 10^5$ in the ambient medium), we have $N \sim 83\,N_{\text{min}}$ and thus meet the requirements imposed by the Jeans condition. Additionally, we would like to emphasize that we also did simulations with $\sim 1/5 $ of the particles and found qualitatively the same behaviour as described here.

\section{RESULTS \& DISCUSSION}
\subsection{Scaling of the magnetic field}

Regarding the evolution of the magnetic field, we show in Fig. (\ref{fig:ncore_bfield}) the dependency of the magnetic field $B$ on the core density $n_c$. Assuming a scaling relation parametrized as $B\propto n^{\kappa}$, as is usually done in the literature \citep[e. g.][]{Heilesuq}, we performed a fit which gave a result of $\kappa = 0.58$. For this fit we considered only the region of the highest density $n_{\text{max}}$ within the collapsing cloud. Increasing the considered density range used for averaging up to $n_{\max}/2 \leq n \leq n_{\max}$, as in \cite{Banerjee2006uq}, we see the same trend only slightly steeper with $\kappa = 0.64$ (not shown). This is in good agreement with \cite{Banerjee2006uq} and \cite{Price2007tg}, who find a value of $\kappa \sim 0.6$, but in disagreement with other studies \citep[e. g.][]{Desch2001fk,Li2004uq,Burzle2011fk}. However, the scaling relation depends strongly on the geometry of the magnetic field.

\begin{figure*}
\begin{center}
\includegraphics[scale=0.57]{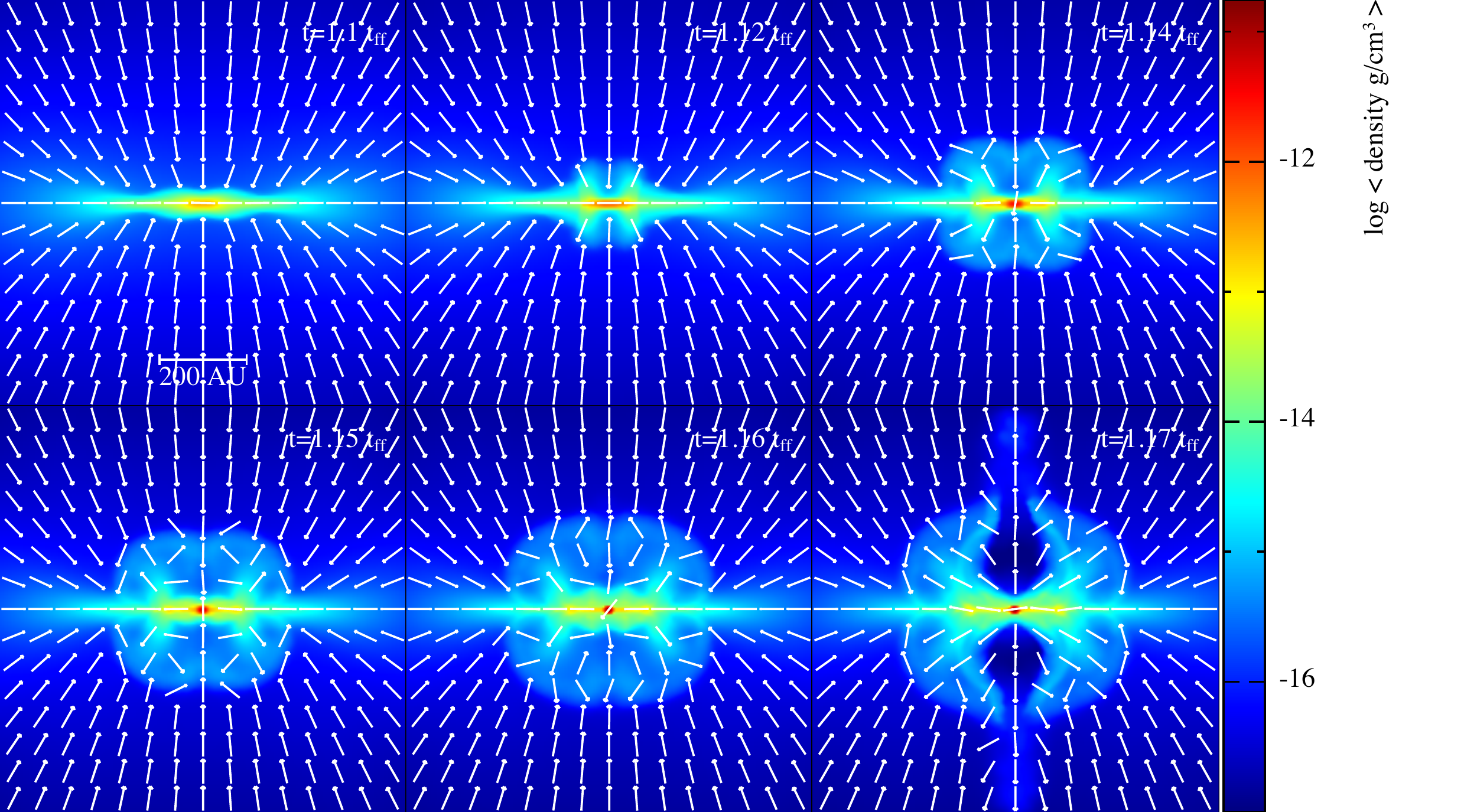}
\caption{\label{fig:rho} The figure shows a sequence of panels with increasing time, starting at $1.10\,\tff$ where the first core forms. The subsequent snapshots were taken at $1.12\,\tff$, $1.14\,\tff$, $1.15\,\tff$, $1.16\,\tff$  and $1.17\,\tff$. Shown is the normalized projection of the density, i. e. an integration along the line of sight divided by the average value, performed in a thin slice around the x-z plane. The arrows which, for better visibility, were set to equal magnitude, indicate the velocity.}
\end{center}
\end{figure*}

\subsection{First core formation and outflow}
We define the adiabatic, or 'first', core as the space occupied by the material exceeding the critical density $\rho_{\textrm{crit}}$, that is where the gas becomes optically thick. We find that the first core has formed at a time of $1.1 \tff$ or $33000\, \unit{years}$. At this time the central density has reached a value of $n_{\max} \simeq 10^{11}\,\unit{cm}^{-3}$, and the mass of the first core is $M \approx 0.02\,\msun$. Due to the fast rotation the first core has an oblate structure, with an equatorial radius of $r_{\textrm{eq}} = 43.5\,\textrm{AU}$ in the $x-y$ plane, and a polar radius of $r_{\textrm{pol}} =9.5\,\textrm{AU}$ parallel to the $z$-axis at formation time. Defining the oblateness as usual, we find a value of $\varepsilon = 1-r_{\textrm{pol}}/r_{\textrm{eq}}=0.78$. 

\begin{figure*}
\begin{center}
\includegraphics[scale=0.57]{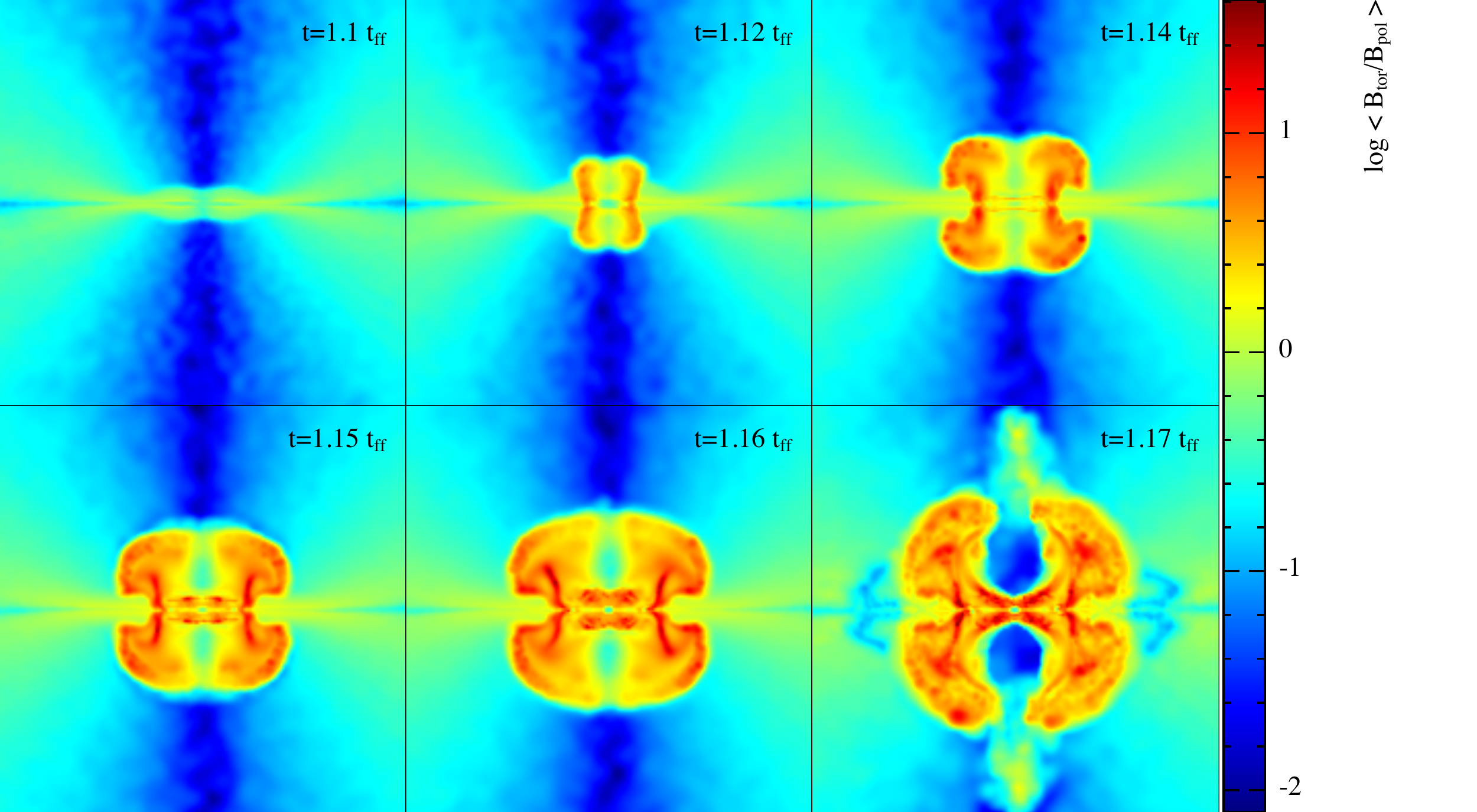}
\caption{\label{fig:btor} The same sequence of panels as in Fig. (\ref{fig:rho}) is shown. Here, the normalized projection of the ratio of the toroidal to the poloidal magnetic field component $B_{\textrm{tor}}/B_{\textrm{pol}}$, is shown. The vectors $\Bbold_{\textrm{tor}}$ and $\Bbold_{\textrm{pol}}$ were inferred from $\Bbold=(B_x, B_y, B_z)$, which is known at every particle position, by expressing the projection of $\Bbold$ onto cylindrical unit vectors in rectangular coordinates. The integration was again performed in a thin slice around the x-z plane.}
\end{center}
\end{figure*}

Only shortly after formation of the first core, at $\sim 1.11 \tff$, a slow bipolar outflow with peak velocities of  $\sim 1.9\,\unit{km}\,\unit{s}^{-1}$ is ejected from the disc. This can be seen in Fig. (\ref{fig:rho}), which shows the normalized density near the x-z plane for different timesteps, and the vectors show the velocity field (arrows have equal magnitude for better visibility).  This outflow is accompanied with the build-up of a strong toroidal magnetic field, as shown in Fig. (\ref{fig:btor}), where the ratio $B_{\text{tor}}/B_{\text{pol}} > 1$ in the outflow regions. This kind of low-velocity outflow, which is frequently found in other simulations \citep[e. g.][]{Tomisaka1998fk,Banerjee2006uq,Hennebelle2008uq,Machida2008fk}, is usually considered as magnetic tower \citep{Lynden-Bell1996fk,Lynden-Bell2003fk,Kato2004uq}. In a magnetic tower, the toroidal component of the magnetic field is continuously generated by the rotating disc and pushes material into the surrounding medium by means of magnetic pressure. Additionally, this can be proven by considering the ratio of thermal to magnetic pressure which, in the outflow region, is given by $\beta_{\text{plasma}} \sim 0.1$ indicating magnetic pressure dominating thermal pressure there.  This magnetic tower flow creates a torus-like structure that further expands into the surrounding gas, also visible in Fig. (\ref{fig:rho}). 

We note that our results suggesting toroidal magnetic pressure as driving mechanism are in good agreement with results presented by \cite{Banerjee2006uq} and \cite{Hennebelle2008uq}, but, however, in disagreement to the findings by \cite{Machida2008fk}. The latter authors attribute the emergence of the slow outflow to magneto-rotational effects rather than magnetic pressure. On the other hand, they had included the effects of non-ideal MHD (ambipolar diffusion and resistivity) which are absent in this work and in \cite{Banerjee2006uq} and \cite{Hennebelle2008uq} which might have an important influence on the magnetic field structure.

Later in our simulation, at $1.17\,\tff$, also a fast outflow, with peak velocities of $\sim 28\,\unit{km}\,\unit{s}^{-1}$, emerges from the central parts of the first core, visible in the last panel of Fig. (\ref{fig:rho}) and Fig. (\ref{fig:btor}), respectively. A similar structure is also visible in \cite{Hennebelle2008uq} in the weak field case, probably also at the same simulation time. However, since we use a comparable set-up as these authors, the same restrictions regarding the thermal structure of the protostar also apply to our work. So, although the velocities found here are comparable to those obtained in other work \citep[e. g.][]{Banerjee2006uq,Machida2008fk}, an identification of this fast flow component with optical jets seen in observations, which believed to be launched during second collapse, seems to be doubtful at this stage. Certainly, further investigations considering a more correct treatment of the thermodynamics within molecular cloud core are necessary. 

\begin{figure}
\begin{flushleft}
\includegraphics[width=\columnwidth]{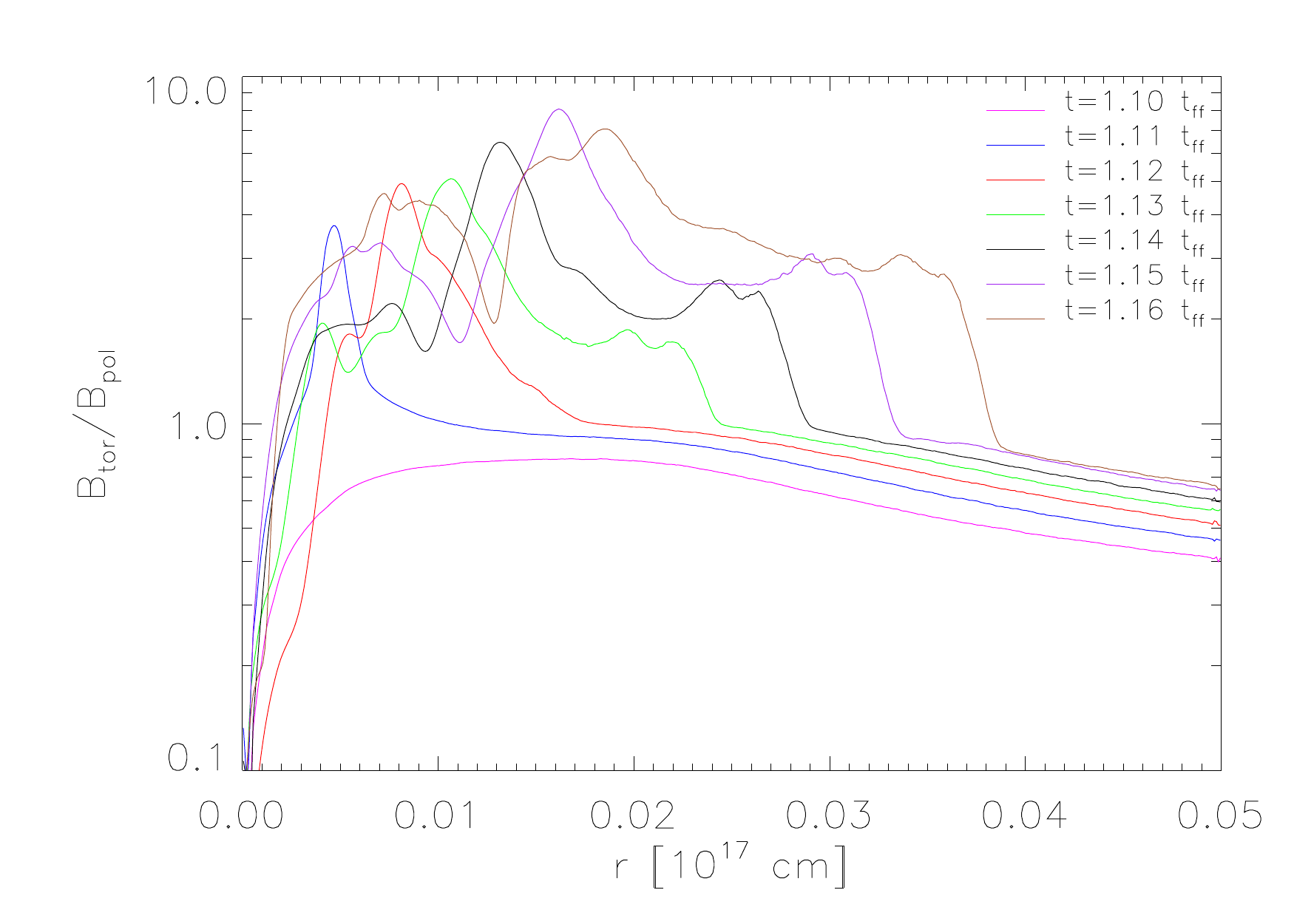}
\caption{\label{fig:btorbpol_plot} Shown is the spacially averaged ratio $B_{\textrm{tor}}/B_{\textrm{pol}}$, plotted against the distance $r$ from the cloud centre for different simulation times. As in Fig. (\ref{fig:btor}), $\Bbold_{\textrm{tor}}$ and $\Bbold_{\textrm{pol}}$ were obtained by expressing the projection of $\Bbold$ onto cylindrical unit vectors in cartesian coordinates. The curves show generation of the magnetic field components and their outward propagation.}
\end{flushleft}
\end{figure}

Finally, in Fig.\ (\ref{fig:btorbpol_plot}) we show a sequence of spatial averages of the ratio  $B_{\textrm{tor}}/B_{\textrm{pol}}$ from $1.10\,\tff$ to $1.16\,\tff$, calculated in the central parts of the cloud core. Here we see the generation of the toroidal part of the magnetic field, as well as its outward propagation with time. However, at $1.15\,\tff$ it is also visible that the toroidal part is re-generated in the very central parts of the disc. Thus, this proves the connection of the outflow with the evolution of the toroidal field component. 

\section{SUMMARY}
In this work, we carried out collapse simulations of magnetised and rotating molecular cloud cores using the Smoothed Particle Magnetodynamics (SPMHD) method.  We reproduced important features of protostellar outflows, qualitatively in good agreement with results reported in the literature. Furthermore, we computed the ratio  $B_{\textrm{tor}}/B_{\textrm{pol}}$ and thus give a quantitative proof of the evolution of the toroidal part of the magnetic field connected with the outflow. 

\section*{ACKNOWLEDGEMENTS}
 Rendered plots were made using the {\small SPLASH} software written by Daniel Price \citep{Price2007hc}. F.B.\ acknowledges granting of computer time from John von Neumann-Institute for Computing (NIC), Jülich, Germany. K.D.\ acknowledges the support by the DFG Priority Programme 1177 and additional support by the DFG Cluster of Excellence 'Origin and Structure of the Universe'. F.S. thanks for the support from DFG Research Unit 1254. 
 R.S.K.\ and P.C.C.\ acknowledge financial support by contract research {\em Internationale Spitzenforschung II} of the Baden-W\"{u}rttemberg Stiftung (grant P-LS-SPII/18) and from the German {\em Bundesministerium f\"{u}r Bildung und Forschung} via the ASTRONET project STAR FORMAT (grant 05A09VHA). R.S.K. furthermore gives thanks for subsidies from the Deutsche Forschungsgemeinschaft (DFG) under grants no KL 1358/10, and KL 1358/11 and via the SFB 881 The Milky Way System. Finally, we would like to thank our anonymous reviewer, whose remarks led to a significant improvement of our paper.
 
\bibliography{buerzle}
\label{lastpage}
\end{document}